\documentstyle[epsf]{kjhantwo}

\title[]{\centering {Diffuse Dark and Bright Objects}
    \\
    {\centering {in the Hubble Deep Field}}
\thanks{Based on observations with the NASA/ESA {\it Hubble Space
Telescope}, obtained at the Space Telescope Science Institute, which
is operated by AURA, under NASA contract NAS 5-26555.}}

\author[Park \& Kim]
{\centerline {Changbom Park and Juhan Kim} \\
 \centerline {Department of Astronomy,
        Seoul National University, Seoul, 151-742 Korea} \\
 \centerline { cbp@astro.snu.ac.kr,
               kjhan@astro.snu.ac.kr}}
\date{\it Received 1997 June 24 ; }

\begin{document}

\maketitle




\begin{abstract}

 In the Hubble Deep Field (HDF) we have identified 
candidate regions where primordial galaxies might be forming. 
These regions are identified from negative or positive peaks 
in the difference maps obtained from the HDF maps
smoothed over $0.8''$ and $4''$.
They have apparent $V$ magnitudes typically between 29 
and 31 (missing flux below the local average level
 for the dark objects), and are much
fainter than the nearby $L_*$ galaxies.  

The identified objects are shown to be real by two ways. First, the 
cross-correlations of these peaks detected in different filters are strong.
The bright objects have the cross-correlation lengths of about $0.3''$.  
Second, their auto-correlation functions
indicate that these faint diffuse objects are self-clustered.   
Furthermore, the auto-correlation function for the high-redshift, 
star-burst subset of bright objects selected by color,
 has an amplitude significantly higher than that of the total sample.  
The subset of objects dark in the F450W and F606W bandpasses, but 
bright in F814W, also shows stronger correlation 
compared to the whole dark sample.
This further supports that our samples are indeed physical objects.
The amplitude and slope of the angular correlation function of the bright
objects indicates that these objects are 
ancestors of the present nearby bright galaxies.  
It is shown that the data reduction artifacts can not be
responsible for our sample.

We have inspected individual bright objects and noted that they have 
several tiny spots embedded in extended backgrounds.  
Their radial light distributions
are diverse and quite different from those of nearby bright galaxies.  
They are likely to be the primordial galaxies at high redshifts
 in the process of active star formation and merging.

The dark objects in general appear smooth.  Our subset of the 
dark objects is thought to be the `intergalactic dark clouds' 
blocking the background far UV light (at the rest frame) at high redshifts
instead of empty spaces between the first galaxies at the edge 
of the universe of galaxies.

\end{abstract}


%
%

\section{Introduction}

The HDF images  (Williams et al. 1996) taken
by the Wide Field Planetary Camera (WFPC-2) on the Hubble Space 
Telescope, have given us the unprecedented opportunity for the study of 
the high redshift universe.  One can easily detect objects with AB 
magnitudes (Oke 1974) down to 27.7, 28.6, 29.0, and 28.4 in the 
F300W (hereafter U), F450W (B), F606W (V), and F814W (I) bandpasses, 
respectively (Madau et al. 1996).  The angular resolution reaches down 
to about $0.05''$.  This enables one to inspect the detailed internal 
structure of faint distant galaxies.
  Therefore, the most important issue one can raise from the HDF data 
is the discovery of proto-galaxies forming at high redshifts, or 
the epoch of galaxy formation.  There have been several studies
on this issue.
  Steidel et al. (1996) has confirmed five high redshift galaxies with 
$(V+I)/2 < 25.3$ which were selected based on the spectral 
discontinuities between the U and B bandpasses. These galaxies are 
found to have compact cores surrounded by diffuse and asymmetric 
halos. Clements and Couch (1996) have also used the presence of the
 Lyman-break in the U filter as a high redshift indicator.
 They have found 8 candidate primeval galaxies with $B<25.7$ that are 
thought to lie between $2.6<z<3.9$.  
These objects are significantly brighter than $L_*$ galaxies, but are 
smaller and irregular than nearby galaxies.     
Lowenthal et al. (1997) have directly measured the redshifts of 
color-selected high redshift candidate galaxies with $(V+I)/2 < 25.5$, 
and confirmed 16 galaxies lying indeed at high redshift $z>2$. 
These galaxies are small but luminous, 
with half-light radii $1.8<r_h<6.5$ h$^{-1}$ kpc, and absolute 
magnitudes $-21.5 > M_B > -23$.  Morphology of the high 
redshift galaxies are diverse, sometimes showing many small knots of 
emission embedded in wispy extended structures.

  Another important problem one can study from the HDF data is the 
evolution of galaxies.  This can be done by looking at numbers, colors,
 and  clustering  amplitude of galaxies as a function of redshift or 
apparent magnitude limit.
Villumsen et al. (1997) has measured the non-zero two-point auto-
correlation functions of  galaxies to a magnitude limit of $V=$ 29.  
They claim that the measured amplitude of angular correlation function 
is consistent with the linearly evolving correlation function of the nearby 
IRAS galaxies.
Colley et al. (1996) has detected an excess two-point angular correlation 
function of faint objects found in the HDF.  The AB magnitudes of most 
of their objects are between 25 and 30 in (V+I)/2.  
They have also found that the high redshift (thought to be at $z>2.4$) 
subset of 695 objects shows much higher correlation strength compared 
to that of the low color-redshift  objects, with the correlation length 
of about $0.9''$.  
Colley et al. (1997) argue that these small compact objects are giant
star-forming regions, whose disks  have been dimmed by K-correction 
and surface brightness dimming.  

As Peebles (1993) and Colley et al. (1997) noted, galaxies in the images
like the HDF are well resolved.  Thus the bright spots in the HDF can 
well be subgalactic units like giant HII regions.  On the other hand, the
 primordial galaxies just starting to form at higher redshift may not have
 giant star-forming regions, but have numerous small starburst clumps 
forming gradual flux excesses over the proto-galaxy scale.  They may
appear even as dark spots blocking the background light.
This expectation is manifest in the Steidel et al. (1996)'s high-redshift 
($2.4<z<3.4$) objects which show diffuse halos with typical diameters 
$ \sim 1.0''$ around very compact bright spots.  
Without the compact cores, they would have only diffuse halos. 

In this paper we find candidate primordial galaxies that are much fainter 
 ($29<V<31$) than those discovered by previous studies, and that appear 
as diffuse ($\sim 1''$) objects.   
We show that these objects are real and physically clustered.  
Finally, we inspect their internal structures to study 
galaxy formation process.  We also pay attention to dark spots in the
HDF images which may be galaxy-scale dark clouds or open spaces 
between primordial galaxies.

\section{Samples}

The candidate proto-galactic objects are found in the following way.
We smooth the second (WF2), third (WF3) and fourth (WF4) 
field frames of the HDF images 
in the F606W filter over the Gaussian FWHM of 4 pixels ($0.16''$) 
(hereafter our smoothing length means the Gaussian FWHM).   
We do not use the first frame of the HDF images taken by the PC1.
The sky level and the 1 $\sigma$ fluctuation are found from the pixel 
values between 20\% $\sim$ 60\% of the whole frame.   
 We then mask  all those pixels whose pixel values exceed 15 times 
the sky fluctuation above the sky level.
The resulting 
mask regions are expanded by 2 pixels in all directions, and regions 
affected by bright stars are manually masked.   This masking effectively 
hides all sources brighter than 29 in V.  
The mask found in the V filter images is used in all bandpasses .  

The masked maps of the original unsmoothed HDF images are 
smoothed over the proto-galaxy scale.   
We choose the galaxy-scale of $0.8''$ (20 pixels)
 based on the study of Steidel et al. (1996) and Colley et al. (1996). 
By smoothing the maps over $0.8''$, we avoid picking up several objects
belonging to a single galaxy, and significantly increase the 
signal-to-noise ratio. 

The difference maps are then made by subtracting between the masked 
B, V, I maps smoothed over $0.8''$ (20 pixels) and $4''$ (100 pixels).   
We obtain nine (three filters in three fields) difference maps 
(we do not use the U filter images). 
We use the difference map to find
 the very faint peaks because the background level and contamination from 
nearby bright sources have to be precisely determined, and  the objects 
concerned are thought to be small, $\ll 4''$.   
The second version HDF images described by Williams et al. (1996) 
in fact seem to have small sky level errors.   
We have calculated the variation of the most negative pixel value 
in the WF3 as a function of the smoothing 
length $R$.  If the negative flux is caused by
the random noise fluctuation below the sky level, it should vary 
as $\propto 1/R$.  However, it varies as
$|f_{min}| = (|f_{min, 4\, pixel}|-f_0)/(R/4) + f_0$ 
where the mean sky level error
$f_0= 8\times 10^{-7},  -6.8\times 10^{-6}$, and $-3.8\times 10^{-6}$ 
for B, V, and I filters, respectively.  Here $ f_{min, 4\, pixel}$ is the
most negative flux when $R=4$ pixel.
Pictures of the raw images also indicate that the sky level is not uniform.
In a difference map the sky level error is automatically removed, 
and the long-range contaminating light cast from local bright sources 
is also subtracted out (see section 4.1 for the flat field errors, and
4.3 for the contaminations by bright stars and galaxies).

We have then looked for peaks in the difference map 
within the region of 
the CCD chip with pixel indices $310<i<1820$ and $380<j<1800$.  
This restriction is made to avoid the edge effects caused by the 
Gaussian smoothing.
We have found 5359 
peaks with heights between $+0.5\sigma_d$ and $+3.5\sigma_d$ 
in all three HDF frames (WF2, WF3 and WF4), and in all three
bandpasses (B, V, and I).  Here $\sigma_d$ is the $rms$ flux fluctuation 
 in the unmasked region of a difference map. 
We assign magnitudes to peaks using the peak heights assuming that 
each peak has the Gaussian profile with the FWHM of $0.8''$.   
Most peaks are fainter than 29 in the AB magnitude.   
In the case of dark spots the AB magnitudes of the missing flux below
the zero (local background) level is measured in the same way.  
We will designate this
magnitude using the superscript $^-$, as in $^-V$.  
We have also generated higher resolution difference images
by subtracting between maps smoothed over $0.16''$ and $4''$
so that the internal structure of individual objects can be inspected
(see Figure 8).

Steidel et al. (1996) has used the spectral curvature criterion
to select objects whose U magnitudes are affected by the Lyman-break, 
thus are at high redshifts.    We use a similar criterion 
\begin{equation}
{\rm [F450W-F606W] > 1.2 + [F606W-F814W]}
\end{equation}
to make a high redshift, star-forming subset.   Figure 1 is the color-color 
diagram of the bright peaks identified in $V$ with $V<31$, but with $B$ and 
$I<32$.  There are 123 objects satisfying this color criterion (black dots).  
The table and pictures of this 
bright subset can be found in Park \& Kim (1997).
\begin{figure}
\vspace{10cm}
\caption{The color-color diagram of the bright objects with $V< 31$,
and $B$ \& $I< 32$.   The black dots satisfy the color criterion
(1).}
\includegraphics{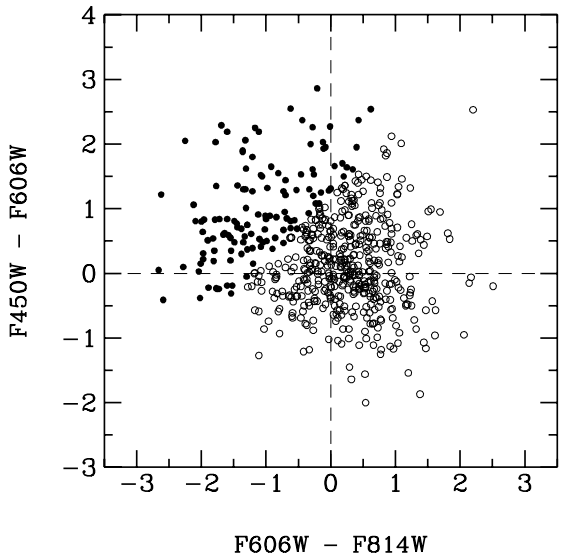}
\label{fig-1map}
\end{figure}

This criterion selects objects blue in [V-I] but red in [B-V].    
Therefore, they are likely to be star-forming blue galaxies 
with their B-bands affected by the redshifted Lyman-break.  
The estimated redshift of our color-selected objects, inferred 
from Figure 1 of Madau et al. (1996), is $z\ge 3.6$.

Assuming the redshift and the spectral energy distribution, we are able
to calculate the absolute magnitude of an object.  Using the HST FOS 
spectra of extragalactic giant star clusters (Rosa \& Benvenuti 1994),
we have calculated $M_V$ corresponding to $V$ of our objects.  
The absolute magnitude of an object with $V=30$, typical in our sample,
is $-14\sim -16$ at $z=3.6$ depending on the slope of the far UV
spectrum.  Here, we used $h=1$, the Hubble constant divided by 
100 km/sec/Mpc. Therefore, the bright objects in our sample are much 
fainter than the present $L_*$ galaxies.

We have inspected the images to check if dark objects are artifacts 
of our background subtraction, and concluded that it is not the case.  
We have grouped the dark objects
into several sub-groups.  One interesting sub-group
consists of 196 dark objects identified in the V filter whose $^-V$ and
$^-B \le 31$, but $I\le 31$.  They are dark in the $B$ and $V$ filters
 but bright in the $I$ filter (see section 3.2 and Figure 8).   
They may be intergalactic dark clouds.   
Spots dark in all three filters may have been caused 
by open dark spaces between first forming galaxies or by dense intergalactic
clouds.  Effects of reduction errors like inaccurate flattening on our 
samples are discussed in section 4.

\section{Correlation Functions of the Selected Objects}

%

\subsection{Cross-Correlation Function} 

If the faint peaks we have found in the HDF are real 
objects, peaks found in one bandpass would in general also appear in 
different bandpasses unless they have very steep colors.  
Since the primordial
galaxies may have large color fluctuations on their surfaces, 
an object in one bandpass may not appear at the same location in 
different bandpass.  We therefore calculate the
cross-correlation function (hereafter CCF) of objects found in one 
bandpass with those in different bandpasses 
instead of looking at the direct pixel-to-pixel correlation.

We measure the CCF from the following formula.
\begin{equation}
w_{C}(\theta) = {N_{dd}\over N_{rr}} {N_{r1}N_{r2}\over N_{d1}N_{d2}}-1,   
\end{equation}
where $N_{d1}$ and $N_{d2}$ are the numbers of objects in two 
bandpasses, 
$N_{r1}$ and $N_{r2}$ are the number of random points we put 
in the sample areas (mask-free region),
and $N_{dd}$ and $N_{rr}$ are the number of pairs of objects and of 
random points at angular separation of $\theta$.
\begin{figure*}
\vspace{10.5cm}
\caption{({\it left}) The cross-correlation functions of the bright objects
 with peak heights greater than 0.5$\sigma_d$ in the $V$
filter, versus those in I or B.  Filled symbols are the
cross-correlation functions between dark objects found in different
bandpasses.  $1\sigma_d$ is the standard
deviation of the flux fluctuation in each bandpass in the mask-free
region of the difference map.
({\it right}) Variation of the amplitude of the cross-correlation
function of the $V$-selected objects with varying peak heights versus
the $I$-selected objects.  The correlation length is about $0.3''$.}
\includegraphics{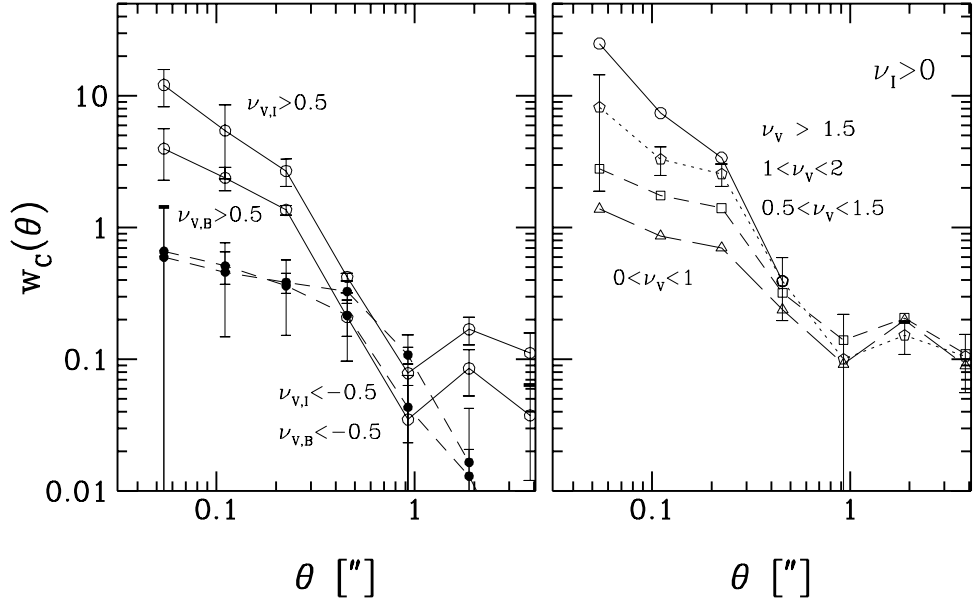}
\label{fig-2}
\end{figure*}

Open circles in Figure 2 ({\it left}) show the CCF of peaks 
with heights greater than 0.5$\sigma_d$ in the $V$ filter 
difference image versus those in $I$ and $V$.  $\sigma_d$ is 
the standard deviation of the background flux fluctuation in the 
mask-free region of the difference map in each bandpass.  
The 1.0, 0.5 and  0.2 $\sigma_d$ 
peaks in $V$ correspond to AB magnitudes of 
about 30.04, 30.78 and 31.73, respectively.  
The uncertainty limits are  obtained from the variation of the CCF over 
the three fields.  Both $w_{VI}$ and $w_{VB}$ show strong correlations
with the correlation length of about $0.3''$.  
Beyond $0.3''$ the CCF drops fast,
and remains small but finite out to $\theta > 4''$, which is due 
to the clustering with different objects.  
The CCF of the $V$ images with the $B$ filter ones is lower. This
is probably due to the lower signal-to-noise ratios of the $B$ images.
 Figure 2 ({\it right}) show the variation of the amplitude of the CCF of the
V-selected peaks with the I-selected peaks.  
As one can expect, higher peaks show 
stronger correlation, but peaks with different heights all have roughly 
the same correlation length of about $0.3''$.  
This means that many peaks seen in one filter also appear 
 as peaks within $0.3''$ in another filter.
Thus, we can conclude that most peaks we have identified are real.
This also justifies our value of the smoothing length 
used to find the peaks as candidate primordial galaxies.

Dark spots show weaker but still significant cross-correlations
across images observed with different filters.  
The black dots in Figure 2 ({\it left}) are their CCFs.  
Therefore, they cannot be mere instrumental artifacts
(refer to section 4 for more discussion).

\subsection{Auto-Correlation Function} 

Once the faint objects we have identified are known to be physical 
objects that are observed in different bandpasses, it is also interesting to 
know if they are self-clustered.   We calculate the two-point 
angular auto-correlation functions (ACF) from the formula 
(cf. Hamilton 1993):

\begin{equation}
w(\theta) = {N_{dd}(\theta) N_{rr}(\theta)\over N_{dr}(\theta)^2} 
{4N_d N_r\over (N_d-1)(N_r-1)}-1.   
\end{equation}

\begin{figure*}
\vspace{9.5cm}
\caption{({\it left}) Auto-correlation functions of bright objects
with the peak heights higher than 0.5$\sigma_d$ in $B$, $V$, and $I$
bandpasses.  Corresponding magnitude limits are 30.6, 30.8 and
30.0, respectively.  At the top is the correlation function of
the subset of the $V$-selected objects with $V < 31$, $B$ \& $I < 32$,
and also satisfying the color criterion inequality (1).
({\it right}) Auto-correlation functions of dark objects with the peak
 heights between  $-0.5\sigma_d$ and $-1.5\sigma_d$.  Also shown at the
top is the correlation function of the subset of the $V$-selected dark
objects with $^-V$ and $^-B$,  magnitudes corresponding
to the missing fluxes, darker than 31,
but with $I$ brighter than 31.}
\includegraphics{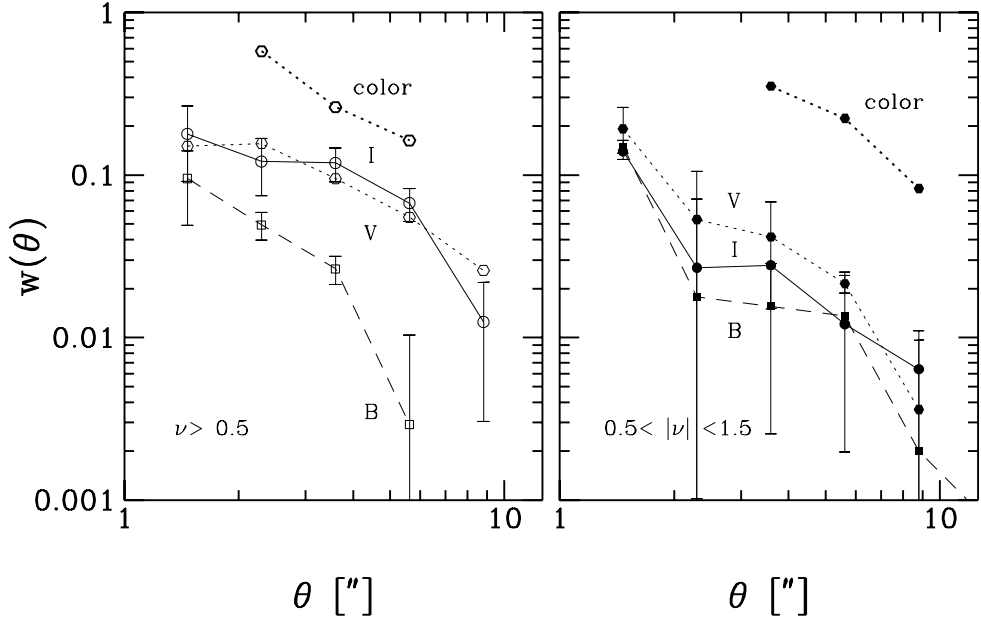}
\label{fig-3map}
\end{figure*}

Here $N_{dd}$ is the number of pairs among $N_d$ objects separated 
by angle $\theta$ in a given field and a given bandpass, 
$N_{dr}$ is the number of pairs between the objects and $N_r$ random 
points.
Figure 3 ({\it left}) shows the angular ACFs of the objects 
whose peak heights are 
greater than $0.5\sigma_d$ in $B$, $V$, and $I$.
These peak heights correspond to the AB magnitudes of about 
30.6, 30.8, and 30.0, respectively.   
The uncertainty limits are again 
estimated from the field-to-field fluctuations. 
 The amplitude of the ACF is low, but definitely positive over the scales
out to $10''$.  The ACF at 
separations $\le 1''$ is dropping because objects have been found 
in maps smoothed over $0.8''$.  
It should be also noted that our estimation of the angular ACF is
lowered by the small size of the survey area.  This is because the angular 
ACF satisfies the integral constraint (Peebles 1980)
\begin{equation}
{1\over \Omega^2}\int\int w(\theta_{12})d\Omega_1 d\Omega_2
 = -{1\over N_d},
\end{equation}
where $\Omega$ is the solid angle of the sample.
In a finite sample the correlation function of objects clustered at small 
scales will be underestimated at scales comparable to the sample size
by the amount given by the above integral.  
If the correlation length is very small compared to the survey size, 
it becomes
\begin{equation}
\Delta w = {1\over \Omega}\int w(\theta)d\Omega.
\end{equation}
When the survey area is a circle with radius $\theta_{\rm max}$, and 
when the angular CF is given by $w = (\theta/\theta_0)^{1-\gamma}$, 
equation (5) gives
$\Delta w = (2/(3-\gamma))(\theta_{\rm max}/\theta_0)^{1-\gamma}$.   
In the case of the WF3 field the effective 
(mask-free) sample area in our analysis is about 3044 arc second$^2$.  
If we approximate it as a circle of radius $31''$, and use $\gamma=2$ 
and $\theta_0=0.2''(1'')$,
 we get $\Delta w\approx 0.01 (0.06)$.  
Therefore, the angular ACFs are not 
actually decreasing as fast as shown in Figure 3.  

If we use the color-selected subset of objects with $V<31$ but with
$B, I<32$, and satisfying inequality (1),
 the amplitude of the ACF increases
significantly as shown by open hexagons in Figure 3 ({\it left}). 
The uncertainty (see Figure 4) is larger due to the small size of 
the subset (123 objects).  
On the other hand, another subset with $V<31$ and $B, I<32$, but satisfying
inequality (1) with 
the direction of the inequality reversed, has a vanishing angular ACF.  
They are those plotted as open circles in Figure 1.
This clearly supports
that the faint objects we selected are real, and physically clustered with 
strengths different for different species.  

Dark objects also show positive ACF up to $\sim 10''$ as shown in
Figure 3 ($right$).
The subset of objects, dark in the short wavelength bandpasses 
$B$ and $V$, but bright in the long wavelength bandpass $I$, 
also shows stronger correlation (filled hexagons in the right 
panel of Figure 3) compared to the whole dark sample.  
On the other hand, we have found nearly zero correlations for the objects 
dark in all $B, V$, and $I$ filters, and also for the objects dark in 
$V$ and $I$ but bright ($B<31$) in $B$ (i.e. blue). 

\begin{figure}
\vspace{10cm}
\caption{Thick curves are auto-correlation functions of the $V$-selected
bright objects and the color-selected subset shown in Figure 3.
The thin solid line is a fit to the Colley et al. (1997)'s
correlation function of their compact objects.
Five lines with labels are the theoretical correlation functions
calculated from the correlation function of the nearby bright galaxies
extrapolated to the sample depth.   Model A, B, C, D, and E correspond
to the cases ($\Omega_0, \epsilon) = (1.0, 0.0), (1.0, -1.2), (1.0,
0.8), (0.4, 0.0), (0.4, -1.2)$, respectively.  $\epsilon$ is a parameter
determining the redshift evolution of clustering (see the text).}
\includegraphics{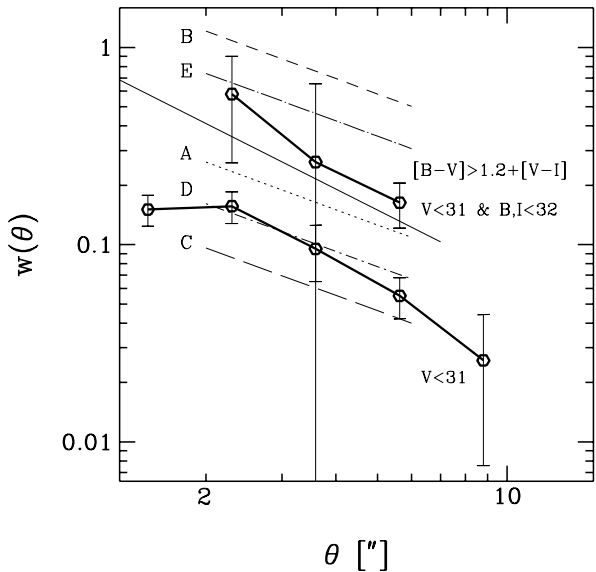}
\label{fig-4map}
\end{figure}

In Figure 4 we have also plotted ACFs of our V-selected bright objects 
and Colley et  al. (1997)'s CF (thin solid line) for a comparison. 
It can be seen that the slope and amplitude of the angular CFs 
of compact brighter spots measured by Colley et al. are close to
that of our diffuse fainter objects at large angular separations 
($\theta > 2''$).
It is interesting that our fainter extended objects are clustered similarly
 with Colley et al.'s brighter and compact objects at large scales.  
Our objects may have less prominent star-forming regions, 
but otherwise be the same as Colley et al.'s sample.

It is interesting to know how the clustering amplitude of these
faint objects compares with that of nearby bright galaxies.   The spatial
two-point CF of nearby galaxies can be approximated by a power-law 
\begin{equation}
\xi(r,z) = (r/r_0)^{-\gamma} (1+z)^{-(3+\epsilon)},
\end{equation}
with $r_0\approx 5.5 h^{-1}$ Mpc and $\gamma\approx 1.8$
 at $r \le 10 h^{-1}$ Mpc. 
We have modeled the redshift evolution of clustering using 
the parameter $\epsilon$ (Peebles 1980).  The case 
$\epsilon = 0$ corresponds to fixed clustering in the proper space
 while $\epsilon=\gamma-3\approx -1.2$ case is when clustering
is fixed in the comoving space.  The CF grows linearly
 in the comoving space when $\epsilon=\gamma-1\approx 0.8$.   

The relation between the spatial CF and the angular
CF at small angular separations ($\theta \ll 1$) is given by the equation
 (Peebles 1980; Efstathiou et al. 1991; Villumsen et al. 1997)
\begin{equation}
w(\theta) = A (r_0/a_0)^{\gamma}{\sqrt \pi} {\Gamma((\gamma-1)/ 2)
 \over \Gamma(\gamma/2)} \theta^{1-\gamma},
\end{equation}
\begin{equation}
A =\int_0^{\infty} g(z) ({d\Sigma\over dz})^2 dz/[\int_0^{\infty} 
({d\Sigma\over dz}) dz]^2,
\end{equation}
\begin{eqnarray}
g(z) &=& \displaystyle x^{1-\gamma}{dz\over dx} (1+z)^{\gamma-\epsilon-3} F(x)
         \nonumber      \\
     &=& \displaystyle x^{1-\gamma}(1+z)^{\gamma-\epsilon-3}H(z) , 
\end{eqnarray}
where $d\Sigma/dz$ is the redshift distribution of the objects under study,
and $H(z)$ is the Hubble parameter.
The function $F(x)$ is defined in the metric
\begin{equation}
ds^2 = dt^2-a^2(dx^2/F(x)^2+x^2 d\Omega^2),
\end{equation}
where $x$ is the coordinate distance at redshift $z$.

Without knowing the function $d\Sigma/dz$ accurately, we use the 
model (cf. Efstathiou et al. 1991)
\begin{equation}
{d\Sigma\over dz} \propto z^2 (\exp[-(z/z_f)^{\beta}]-
\exp [-(z/z_b)^{\beta}]).
\end{equation}
Since we have removed foreground bright objects with $V< 29$,
 and selected objects brighter than 31 in V, 
the mean redshifts $z_b$ and $z_f$ correspond to the magnitude limits 
of 29 and 31.  According to Table 1 of Villumsen et al. (1997), 
they can be estimated
for the HDF as $z_f=2.16$, $z_b=1.87$, and $\beta=2.5$.  
Altering these parameters by significant factors do not change our 
conclusions.  By using this model for the redshift distribution 
of our faint objects, we can calculate their angular CF. 
 In Figure 4 we have plotted theoretical angular CFs for 
five models with zero cosmological constant: 
the density parameter $\Omega_0 = 1.0$ models with
 $\epsilon=0$(A), $-1.2$(B), 0.8(C), and $\Omega_0=0.4$ models 
with $\epsilon=0$(D), and $-1.2$(E). 
One can learn from this figure that the amplitude of angular CF of
our primordial galaxy candidates, extrapolated in various evolution 
and universe models, is of the same order of that of 
nearby bright galaxies.  The ACFs of our and Colley et
al.(1997)'s high redshift subsets are more consistent with negative
$\epsilon$ evolution models, in particular for the open universes.
It implies that the clustering is more or less constant in the comoving
space, and that the objects are statistically rare peaks in the matter 
distribution.

\section{Effects of Instrumental Artifacts}

Since we have selected very faint objects from the HDF images, 
it has to be carefully examined if instrumental artifacts 
are responsible for our sample.
We have estimated the effects of three potentially important
instrumental artifacts on our results.  They are flat field errors, 
effects of dithered hot and cold pixels, and effects of the point
spread function (PSF) near bright stars and galaxies.

\subsection{Flat Field Errors}

The flat field calibration files used in HDF data reduction are 
old, and new superior flat field calibration files
 with corrections on large scales are released. 
We take the difference between the old and new flats of $800\times 800$
sizes, and make flat-difference files of $2048\times 2048$ size by
dithering the frames within roughly $2.3''$. 
Here we have followed exactly the same reduction process used for the HDF
(Williams et al. 1996).
We then smooth these flat-differences over $0.8''$ and $4''$ and subtract
between them.  The resulting files are the errors in the original flats 
with respect to the new flats transferred to our difference images
from which our samples were selected.
\begin{figure}
\vspace{11.6cm}
\caption{Maximum amplitudes of spurious peaks caused by
hot pixels as a function of the height of hot pixels.
Unidentified hot or cold pixels can produce many false peaks
in drizzled images.
The scatter is due to the pixel geometry distortion in
WFPC2 CCD chips (Holtzman et al. 1995).  It can be seen that
the maximum flux contribution by pixel defects (shown as
horizontal dashed lines in Figure 5) is much lower than
the typical amplitude of our bright and dark objects.}
\includegraphics{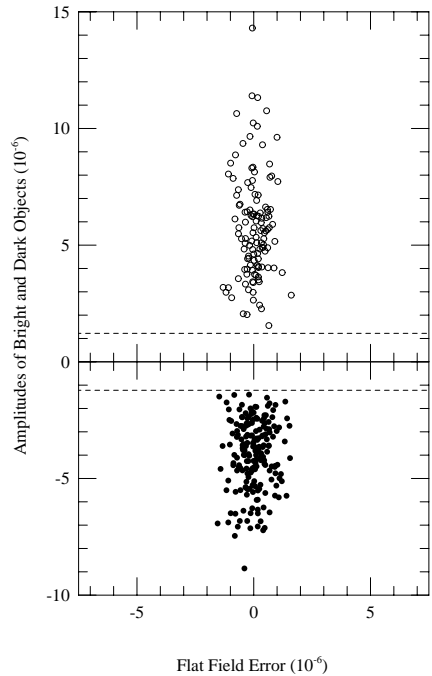}
\label{fig-5map}
\end{figure}

In Figure 5 we show the flat field errors at the locations of our 
color-selected bright and dark objects.
The typical flat field error is about $0.6\times 10^{-6}$ in flux 
while the amplitude of our peaks is typically $5\times 10^{-6}$.
It is clear that the flat field errors can not be responsible for
our bright and dark objects.

\subsection{Hot and Cold Pixels}

In the HDF data reduction process, most hot and cold pixel defects
are identified as those above 5 times the background
sky fluctuation ($\sigma$) on images median filtered  over
 $5\times 5$ pixels, and then masked (the growing-pixel mask;
see section 4.3 of Williams et al. 1996). 
Under such a criterion, hot or cold pixels with peak heights
below $5\sigma$ can not be found.  
These unidentified hot or cold pixels can appear in the drizzled
images as groups of bright or dark objects in our difference maps.
For example, in the $V$ bandpass each unidentified hot pixel will be
dithered to form 11 peaks within roughly $2.3''$ on the final
$2048\times 2048$ frame.  If a significant fraction of our sample
is produced by these pixel defects, they will cause false angular
clustering over that scale.

\begin{figure}
\vspace{7.8cm}
\caption{Maximum amplitudes of spurious peaks caused by
hot pixels as a function of the height of hot pixels.
Unidentified hot or cold pixels can produce many false peaks
in drizzled images.
The scatter is due to the pixel geometry distortion in
WFPC2 CCD chips (Holtzman et al. 1995).  It can be seen that
the maximum flux contribution by pixel defects (shown as
horizontal dashed lines in Figure 5) is much lower than
the typical amplitude of our bright and dark objects.}
\includegraphics{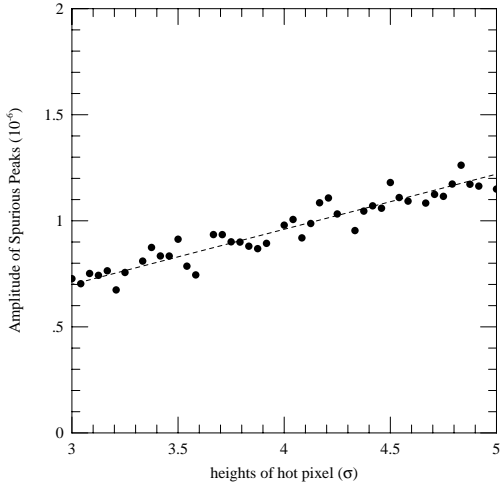}
\label{fig-6map}
\end{figure}

To estimate the amplitude of hot or cold pixels in our final difference
images we have put artificial hot pixels with heights less than $5\sigma$
on a $800\times 800$ array and make the corresponding high
resolution map by projecting the dithered images on
a $2048\times 2048$ array.  
This map is smoothed over $0.8''$ and $4''$ 
and subtracted from each other to make the difference map.  
We then find the maximum peak
out of the spurious peaks caused by the hot pixels.
Figure 6 shows the relation between the height of input artificial
hot pixel and the maximum amplitude of resulting peaks on the
difference map.
It shows that the maximum contribution of 5 (2.5) $\sigma$ hot or 
cold pixels to the flux is only 1.2 (0.6) $\times 10^{-6}$ 
in the difference map.
The horizontal dashed lines in Figure 5 indicate this level.
It is noted that all our bright and dark objects have amplitudes higher
than this maximal effect of unmasked hot or cold pixels.

The flat field error and the hot or cold pixels cannot be responsible 
particularly for the subset of our dark objects mentioned above
since they have negative fluxes on the difference maps
in B and V, but have positive fluxes in I.
Similarly, the color-selected subset of the bright objects are 
unlikely to be affected by these artifacts.

\begin{figure}
\vspace{11.6cm}
\caption{Contamination by long-range wings of the point spread function
at the locations of our color-selected bright and dark objects.
Amplitudes of all of our objects are significantly higher than the
flux leakages from the masked bright stars and galaxies.}
\includegraphics{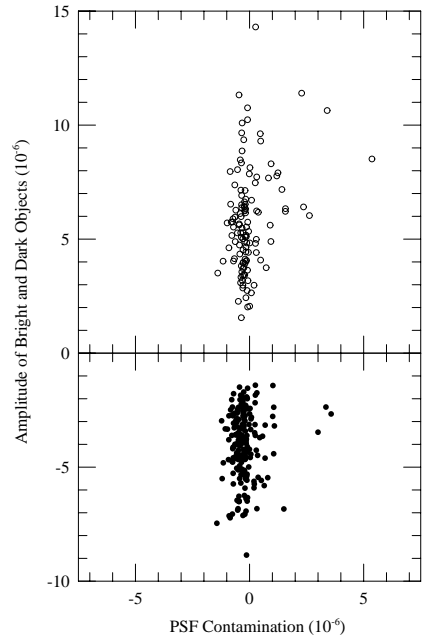}
\label{fig-7map}
\end{figure}

\subsection{Contamination by Bright Stars and Galaxies}

We have masked objects brighter than roughly $V=29$ 
on the HDF images to search for fainter objects.  However,
the PSF of the HDF images is still significant
even out to a few arc-seconds.   Thus bright stars and 
galaxies in the HDF images can cast light out of the masked regions
and might cause false peaks and correlations in our difference maps.

To see the effects of these wings we use the PSFs calculated by
the software Tiny Tim V4.3 recommended by the WFPC2 
Instrument Handbook (Biretta et al. 1996).  
This PSF has a maximum radius of $7.5''$.  
Using this PSF, we have deconvolved the original HDF data to get 
unblurred images.   We then take those pixels which have been masked in
our analysis.   These pixels contain stars and galaxies brighter
than about $V=29$.   The map with these bright sources present but with
zero flux outside them is prepared, and convolved with the PSF.   
This blurred image is smoothed over $0.8''$ and $4''$ and subtracted from 
each other to make the difference map.     From the resulting
map we can measure the effects of bright sources on the flux
distribution outside our mask regions.  We have then measured the 
flux leakages from bright sources at the locations of our bright
and dark objects, and plotted them in Figure 7.   It is again clear
that our samples and correspondingly their correlations are not 
artifacts caused by long-range wings of bright stars and galaxies.
In particular, large contaminations ($\sim 3\times 10^{-6}$) 
on the dark objects are all positive, and these objects 
would have appeared even darker without the contamination.

\section{Morphology  of Peaks}
In the previous section we have shown that the faint diffuse objects 
we have identified are clustered with an amplitude consistent with 
that of the present bright galaxies.  
It supports the idea that these objects are indeed the ancestors of 
the nearby bright galaxies, undergoing formation at high redshifts.
Therefore it is necessary to inspect their morphology 
to understand the galaxy formation.

\begin{figure*}
\vspace{21cm}
\caption{Images of typical bright (top three rows) and dark
(bottom three rows) objects in WF3 detected in $V$
(see the text for more description).
Each panel is a part of the difference map between the WF3 images
smoothed over $0.16''$ and $4''$, and has a field of view of
$3.24''\times 3.24''$.
Objects are at the center, and have sizes of $\sim 0.8''$.
Bright objects are from the color-selected subset.
Dark objects are those with $^-V$ \& $^-B < 31$ but with $I< 31$.  That
is, they are dark in $B$ and $V$ but bright in $I$.  Surface brightness
or darkness profiles in $V$ are also shown.}
\includegraphics{f8.eps}
\label{fig-5map}
\end{figure*}

In WF3, for example, we have found 
40 bright objects with $V< 31$ and $B, I < 32$, and also 
satisfying the color-selection criterion.
Among these the upper three rows of Figure 8 show 
high-resoluton $B, V,$ and $I$ images 
of the 8th (top row, $V=$29.37), 13th (29.45), 18th (29.61) 
brightest objects.
These are from the difference map between the images smoothed
over $0.16''$ (4 pixels) and $4''$ (100 pixels).
  Each panel has a field of view of $3.24''\times 3.24''$, 
and contains one object at the center about four times smaller in length.  
Figure 8 also shows the surface brightness profile in $V$.

Several common features can be noticed.  
First, most of these objects do not have bright centers or dominating 
cores.  They are sprinkled with many noise-like glares and 
show extended backgrounds.  Some of them are highly elongated with 
emission of connections and seem to be undergoing merging.   
When the local background is further subtracted out 
within a bright object, each small spot contained in the object
typically has the $V$ magnitude of $31 \sim 33$.
If they are real signals, they are as bright as the bright
superassociations which can be found in nearby late type galaxies.
Wray \& de Vaucouleurs (1980) have found ultrabright blue 
superassociations in spiral and irregular galaxies having absolute 
$M_B$ magnitudes  between $-10$ and $-15$.   
If these superassociations have the spectra of nearby extragalactic 
giant star clusters (Rosa \& Benvenuti 1994), 
they would appear as objects with $V=29\sim 34$  
(for $h=0.8$) at $z=3.6$, the lower bound of 
redshift of our color-selected subset.  Therefore, 
the noisy spots embedded in an object have brightness of the
superassociations of young blue stars.

Colley et al. (1997) have argued that the objects they have found in the
HDF are probably giant star-forming regions. 
Their argument is based on the comparison with the luminosity, size, 
surface texture and brightness
 profile of nearby giant HII regions like 30 Dorados in LMC. 
Nearby superassociations including 30 Dorados would have 
$V$ magnitudes fainter than $28.5$ at their assumed redshift 
$z>2.4$.   Since their objects are as bright as $V=25$, their objects 
are several magnitudes brighter than the superassociations 
mentioned above.
If their objects are giant star-forming regions, the star formation 
activity in such spots is an order of magnitude stronger than 
nearby super-starclusters.

The surface profiles of our proto-galaxy candidates are diverse, 
and do not resemble 
those of the present bright galaxies.  Color fluctuations on the surfaces 
of the objects are strong.  The brightest region in an object often
appears at a slightly different location in different bandpasses.
Because of our color criterion the color-selected subset of the bright
objects are in general fainter in B and I than in V.  
However, there are often flux excesses in the B and I 
bandpasses nearby the V-selected peaks.    All these characteristics 
support that these objects are in the process of formation.  

The three bottom rows show $0.16''$ resolution $B, V$ and $I$ images of 
three typical objects that are dark in $V$ and $B$, but bright in $I$.
  In $I$ images they are associated with many emission knots. 
Their surface `darkness profiles' in $V$ are also diverse 
like the bright objects.
The dark objects are not highly elongated `lanes'.  This subset of dark 
objects might be the `intergalactic dark clouds' blocking 
the background short wavelength light instead of empty dark spaces 
between galaxies at highest redshifts.

%
%

\section{Conclusions}

We have found very faint ($29<V<31$) extended ($\sim 1''$) bright and
dark objects in the HDF,
 and shown that they are real and physically clustered.  
 We have been able to detect these
objects by removing all sources brighter than about $V=29$ in the HDF
images, smoothing the maps over $0.8''$ and $4''$,
 and subtracting between them to remove the local background.
With significantly increased 
signal-to-noise ratios even $0.5\sigma_d$ peaks in the difference maps 
show strong cross-correlations between images in different bandpasses.  

Taking into account the amplitude and shape of the angular 
auto-correlation functions, the morphology, color and surface 
brightness profiles of these objects, we conclude that they are 
likely to be the primordial galaxies in the process of formation,
 and ancestors of the present bright nearby galaxies.  
The angular CF has a slope of about $-1$, consistent with
the slope of the spatial and angular CFs of the present bright galaxies. 
Its amplitude is also consistent with that of the spatial CF 
of nearby galaxies extrapolated under various scenarios 
for the evolution of clustering, and in various cosmologies.   
More accurately determined CFs of deep objects will be able to 
discriminate these scenarios and cosmological models.  
The subset of high redshift objects constrained by colors 
shows an angular CF significantly stronger than that of
the whole sample.  These objects are typically sprinkled with many
 tiny bright knots surrounded by diffuse backgrounds, lacking the 
predominant core.  Their surface brightness profiles are diverse 
and are far from those of dynamically relaxed objects.      
They are likely to be in the process of formation.

Recent cosmological simulations of structure formation (Park 1997) 
have shown that the galactic and sub-galactic scale objects 
form (undergo complete collapse) at redshifts much before 10 
in popular cosmogonies like 
the standard Cold Dark Matter model (Park 1990; Park et al. 1994).
These high redshift objects continue to grow through frequent
mergings.  Therefore, it is natural to expect to find actively
star forming irregular proto-galactic objects in deep images.
Our proto-galaxy candidates have many of the characteristics
(magnitude, color, morphology, etc.)
of these objects seen in the simulations.
More deep ground and space infrared images are also predicted to show 
these objects (cf.  Djorgovski et al. 1995).  

We have also payed attention to dark patches with negative fluxes below
the local background level.  
Since they tend to persist as the smoothing length is 
increased, and often have dark pairs at short distances in different 
bandpasses, they are thought to be real objects.  
In his deep surveys from 0.32 to 0.9$\mu m$ Tyson (1990) has also noted 
the presence of reproducible  `dark lanes' at scales smaller than $30''$. 
He has argued that they may be intergalactic dust clouds or open
tunnels in the galaxy distribution.  Our dark objects are found to be weakly self-clustered.   The subset of dark objects selected in the
$V$ images, which are also dark in $B$ but bright in $I$ (i.e. red),
do have a significantly higher auto-correlation function compared to
that of the whole dark sample.  On the
other hand, the objects dark in all filters have a nearly zero CF.  The
objects dark in $V$ and $I$  but bright in $B$ (i.e. blue) 
similarly have a vanishing CF.  The first subset of dark objects may be
intergalactic dark clouds opaque in the far UV (in the rest frame), but 
emitting or transmitting some light at longer wavelengths.  
This kind of objects might
supply us an important clue to structure formation mechanisms.
The second subset of dark objects may be
 the empty dark spaces between primeval galaxies.
If this is the case, we are now looking at the edge of the universe
of galaxies.

\bigskip\bigskip\noindent
ACKNOWLEDGEMENTS

This work was supported by the Basic Science Research Institute 
Program, Ministry of Education 1995 (BSRI-97-5408).

\end{document}